# Titania may produce abiotic oxygen atmospheres on habitable exoplanets

***Norio Narita[1,2,3], Takafumi Enomoto[3,4], Shigeyuki Masaoka[3,4], Nobuhiko Kusakabe[2]**

[1]Astrobiology Center, National Institutes of Natural Sciences, 2-21-1 Osawa, Mitaka, Tokyo 181-8588, Japan

[2]National Astronomical Observatory of Japan, 2-21-1 Osawa, Mitaka, Tokyo 181-8588, Japan

[3]SOKENDAI (The Graduate University for Advanced Studies), Shonan Village, Hayama, Kanagawa 240-0193, Japan

[4]Institute for Molecular Science, 5-1 Higashiyama, Myodaiji, Okazaki, Aichi 444-8787, Japan

Corresponding author: Norio Narita (norio.narita@nao.ac.jp)

## Abstract

The search for habitable exoplanets in the Universe is actively ongoing in the field of astronomy. The biggest future milestone is to determine whether life exists on such habitable exoplanets. In that context, oxygen in the atmosphere has been considered strong evidence for the presence of photosynthetic organisms. In this paper, we show that a previously unconsidered photochemical mechanism by titanium(IV) oxide (titania) can produce abiotic oxygen from liquid water under near ultraviolet (NUV) lights on the surface of exoplanets. Titania works as a photocatalyst to dissociate liquid water in this process. This mechanism offers a different source of a possibility of abiotic oxygen in atmospheres of exoplanets from previously considered photodissociation of water vapor in upper atmospheres by extreme ultraviolet (XUV) light. Our order-of-magnitude estimation shows that possible amounts of oxygen produced by this abiotic mechanism can be comparable with or even more than that in the atmosphere of the current Earth, depending on the amount of active surface area for this mechanism. We conclude that titania may act as a potential source of false signs of life on habitable exoplanets.

## Introduction

In recent years, the search for and characterization of extrasolar planets (exoplanets) has become one of the most active research fields in astronomy. NASA's Kepler mission has discovered a number of habitable exoplanets, including a nearly Earth-sized planet Kepler-186f[1]. In 2017, NASA's next mission, Transiting Exoplanet Survey Satellite (TESS), will be launched to find habitable Earth-like planets around stars in the vicinity of our Solar System[2]. The next biggest milestone would be detecting biomarkers on such habitable exoplanets. Then what can we consider as certain evidence of life on distant habitable exoplanets? Previously, the existence of oxygen in the atmosphere has been proposed as a promising biomarker[3,4]. Indeed, next-generation extremely large telescopes (ELTs) are considered to be able to detect atmospheric oxygen on nearby habitable exoplanets, at least in principle[5,6]. However, some studies recently pointed out that abiotic oxygen might build up on habitable exoplanets especially around M dwarfs, due to photodissociation of water molecules in upper atmospheres by extreme ultraviolet (XUV) light[7-9]. Those studies are based on the facts that habitable exoplanets around M dwarfs are orbiting very close to their host stars and such planets receive relatively higher XUV fluxes than does Earth[7]. The studies have raised the possibility that atmospheric oxygen may become a false positive of life on habitable exoplanets.

On the other hand, previous studies have only considered photodissociation of water molecules in upper atmospheres where XUV irradiation from host stars is significantly strong. Such photodissociation is only effective if the cold trap mechanism of water vapor is not efficient and water vapor is abundant in the upper atmosphere[8,10]. Thus oxygen accumulation by XUV photodissociation may be limited if water molecules are only trapped on the surfaces of planets (as ocean) or in the lower atmosphere (as water vapor) where strong XUV light is completely blocked by planetary atmospheres.

In this paper, we propose a new hypothesis for oxygen build-up on habitable exoplanets. We find that the photocatalytic effect of titanium(IV) oxide (hereafter, titania) can contribute to the photodissociation of water on the surface of habitable exoplanets using near ultraviolet (NUV) light (280 − 400 nm) that can reach planetary surfaces. Our estimation shown below indicates that a substantial amount of abiotic oxygen can be accumulated without strong XUV irradiation. This mechanism can occur not only on habitable exoplanets around M dwarfs but also on such planets around FGK-type stars. We thus conclude that titania may act as a potential source of false signs of life on habitable exoplanets.

## Results

・ Properties of Titania

Titania is a naturally occurring mineral with a chemical formula of $TiO_2$. In the Universe, titania forms as dusts in outflow around asymptotic giant branch (AGB) stars and supernovae, and is known to be abundant in meteorites and the Moon in our Solar System. Thus titania is considered to be also common in exoplanetary systems[11].

In addition to its astrogeological importance, titania is known as a photocatalyst in the oxidation of water into molecular oxygen under NUV irradiation[12-14]. In the photocatalytic process (Figure 1), titania absorbs a photon with a wavelength shorter than 400 nm to afford a pair of positive hole ($h^+$) and negative electron ($e^-$) on the titania photocatalyst (Eq. (1)). In the presence of water and an appropriate electron acceptor, the hole oxidizes water to form oxygen (Eq. (2)) and the electron reduces the acceptor (Eq. (3)).

$$\text{NUV photon} \rightarrow e^- + h^+ \text{ (photoexcitation of photocatalyst)} \quad (1)$$

$$\tfrac{1}{2}H_2O + h^+ \rightarrow \tfrac{1}{4}O_2 + H^+ \quad (2)$$

$$\text{Acceptor} + e^- \rightarrow \text{Acceptor}^- \quad (3)$$

It is known that many molecules and ions can serve as electron acceptors in the above photocatalytic scheme[15-17]. Such ions would naturally exist in oceans on habitable exoplanets. Recent experimental studies on photocatalytic properties of titania have shown that the quantum efficiency of the above photocatalytic scheme (Eqs. (1-3)) is as high as 10 percent under abundant electron acceptors[15,17]. It means that one NUV photon can produce 2.5 percent $O_2$ molecule according to Eqs. (1-2).

・ NUV Irradiation on the Surface of the Earth

We investigate NUV irradiation on the surface of the Earth using data taken by the Monitoring Network for Ultraviolet Radiation, started in 2000 and operated by the Center for Global Environmental Research (CGER), Natural Institute for Environmental Studies (NIES). To duplicate a situation that solar fluxes are irradiated from the zenith (namely, NUV irradiation for the subsolar point), we have obtained NUV and total solar flux density data from the Hateruma Observatory at N: $24^{o}03'14''$, E: $123^{o}48'39''$, several meters above sea level. Figure 2 plots examples of NUV (280 − 400 nm) and total solar flux densities on clear and cloudy days around the summer solstice. The data show that total solar NUV lights irradiated from the zenith can reach about 60 W $m^{-2}$ on clear days and above 10 W $m^{-2}$ even on cloudy days. The level of NUV flux density values does not largely change year to year from 2000 to 2014.

・ Order-of-Magnitude Estimation for Abiotic Oxygen Produced by Titania

Here, we make an order-of-magnitude estimation for the possible amount of oxygen produced on the Earth by the photocatalytic process of titania based on the above NUV fluxes. As shown in the Method section, if we assume the quantum efficiency of the above photocatalytic scheme (Eqs. (1-3)) is 10%, then NUV flux density of 1 W $m^2$ can produce oxygen at the rate of about 76 g $m^{-2}$ $yr^{-1}$. Earth's effective area for solar irradiation is $\pi R_{Earth}^2$ = 1.3*$10^{14}$ $m^2$. Here we define the mean surface area ratio where the photocatalytic process of titania can occur (hereafter, titania-active area) as $f_{titania}$. We can neglect the effect of Earth's rotation since we have defined $f_{titania}$ as a mean value. We can also neglect the effect of additional atmospheric extinction due to airmass for the area away from the subsolar point, because the effect has less impact on the order-of-magnitude estimation. As a result, NUV flux density of 1 W $m^{-2}$ for the Earth can produce ~1 * $10^{16}$ * $f_{titania}$ g $yr^{-1}$, corresponding to ~6 * $10^{17}$ * $f_{titania}$ g $yr^{-1}$ for a case of the current NUV irradiation (60 W $m^{-2}$).

- Constraint on $f_{titania}$ for the Earth

Let us consider what we can say from the above order-of-magnitude estimation.

Given that the current NUV flux density (~60 W $m^{-2}$) is long-lasting and the whole surface area is titania-active ($f_{titania}$=1), then such NUV irradiation can potentially dissociate the amount of water in the Earth's ocean ($1.4*10^{24}$ g) in about $2*10^7$ yr. This fact suggests $f_{titania}$<<0.5% for the Earth, since the Earth's ocean has not run dry. On the other hand, the same assumptions predict that the amount of oxygen in the current Earth's atmosphere (~$1*10^{21}$g) can be generated in about $2*10^4$ yr. As such a significant amount of oxygen production due to titania did not occur in the history of the Earth (over $4*10^9$ yr), we can put a very stringent constraint of $f_{titania}$<<$5*10^{-7}$ for the Earth. More specifically, it means the effective titania-active surface area on the Earth is much less than ~250 $km^2$.

・ Possible Abiotic Oxygen on Habitable Exoplanets around Sun-like Stars

Although the stringent constraint seems to be fulfilled at least for the Earth, the above fact means that the photocatalytic process of titania is capable of producing a comparable amount of oxygen in the current Earth's atmosphere on Earth-twin planets around stars similar to the Sun, if such Earth-twin planets have sufficient titania-active areas. The above result implies that titania can potentially dissociate sufficient surface liquid water to produce amounts of oxygen similar to or greater than those in atmospheres of Earth-twin planets around Sun-like (G2-type) stars over time. It means that abiotic oxygen can be generated on habitable planets around Sun-like stars through the photocatalytic process of titania.

・ Cases for Habitable Exoplanets around Other Stellar Types

Let us consider further cases for habitable exoplanets around other stellar types. Hereafter we assume that Earth-twin habitable exoplanets are orbiting around stars of various types at an orbital distance with an effective stellar flux equal to that for the Sun-Earth system (namely, $S_{eff}$ = 1). Given that Earth-twin planets have the same atmospheric scattering/extinction properties as the Earth, we can estimate irradiated NUV flux on surfaces of such planets relative to that on the Earth ($NUV_{ratio}$) by the following equation:

$$NUV_{ratio} = NUV_{top} / NUV_{top,Earth} \quad (4)$$

Here $NUV_{top}$ and $NUV_{top,Earth}$ are NUV fluxes at the top of atmospheres of the Earth and Earth-twin habitable exoplanets around stars of any type, respectively.

Using the procedure described in the Method section, we estimate $NUV_{ratio}$ for Earth-twin habitable exoplanets around host stars of M6, M0, K2, and F6 types. We also compute a fiducial case ($NUV_{top,Earth}$) using stellar parameters for a G2-type star like the Sun. Assumed stellar parameters, planetary orbital distance, and resultant $NUV_{ratio}$ are summarized in Table 1. We roughly estimate necessary $f_{titania}$ ($f_{titania,1Gyr}$) and corresponding effective surface area ($A_{titania,1Gyr}$) to produce the amount of oxygen in the current Earth's atmosphere (~$1*10^{21}$g) in 1Gyr ($10^9$ yr). We also present those values in Table 1. As a result, $f_{titania,1Gyr}$ for all cases are well below 1, meaning that a significant amount of abiotic oxygen can be potentially produced on any Earth-twin habitable exoplanets around various types of host stars within 1 Gyr as long as sufficiently large titania-active areas exist on their surfaces.

・Prerequisites and Limitations of Titania Photocatalytic Reactions

We have presented that the photocatalytic mechanism of titania has a potential to produce abiotic oxygen on habitable exoplanets around various types of stars. This mechanism occurs, however, only if four components in Figure 1 (liquid water, electron accepters, titania, NUV photon) present together. Such environments may exist on habitable exoplanets having oceans (liquid water) with volcanic activities, which can supply oceans with abundant electron accepters. Titania can be supplied onto planetary surfaces via meteorite impacts, although the amount of surface titania would vary from planet to planet. Sufficient NUV photons can reach surfaces of Earth-twin habitable exoplanets. Although thick clouds would reduce NUV irradiations on surfaces by an order-of-magnitude as shown in Figure 2, the current mechanism still works. The proposed mechanism would most efficiently work on habitable exoplanets with shallow liquid water (shallow places or wetlands), but would not work if deep oceans totally surround planets. The current mechanism would halt if electron acceptors deplete in oceans due to either short supply or consumption by other chemical reactions. Thus a stable redox cycle is essential for titania photocatalysis to cause long-term oxygen productions.

・Feasibility of $O_2$ Accumulation in Planetary Atmospheres via Titania Photocatalytic Reactions

Although the photocatalytic reactions of titania can potentially produce abiotic oxygen, it does not immediately imply that oxygen can accumulate in planetary atmospheres. To cause planetary oxidation like the Great Oxidation Event (GOE) that occurred on the Earth via titania photocatalysis, its oxygen flux needs to overtake all the oxygen sinks in the ocean, seafloor, and atmosphere. A major oxygen sink would be $Fe^{2+}$ in the ocean as was the case on the Earth. For the case of the Earth, it is said that cyanobacterial photosynthesis formed large-scale iron depositions (known as banded iron formation, or BIF) by massive oxygen productions with the rate of $\sim 10^{12}$ mol $yr^{-1}$ during the Late Archean[18,19]. This rate can be a lower limit of the oxygen production rate necessary to cause planetary oxidation.

In addition, continuous oxygen productions would be required to maintain oxygen atmospheres even after planetary oxidation. As for the modern Earth, a large amount of oxygen ($\sim 2*10^{13}$ mol $yr^{-1}$) is lost due to various reasons, such as continental oxidative weathering and seafloor oxidation, but the amount is still balanced with that produced from oxygen sources, such as organic carbon burial ($\sim 1*10^{13}$ mol $yr^{-1}$) and pyrite burial[20]. As the organic carbon comes from the activity of photosynthetic organisms, to maintain the oxygen atmospheres in abiotic environments, titania photocatalytic reactions must compensate oxygen for the organic carbon burial.

In summary, an oxygen production rate of over $\sim 1*10^{13}$ mol $yr^{-1}$ would be necessary to cause planetary oxidation and to maintain oxygen atmosphere. Titania photocatalysis can produce oxygen at the rate of about 2.4 mol $m^{-2}$ $yr^{-1}$ under the NUV flux density of 1 W $m^{-2}$. Given that the current NUV flux density on the Earth (~60 W $m^{-2}$), it means that the effective titania active area of $f_{titania} \sim 5*10^{-4}$ ($A_{titania} \sim 7*10^{4}$ $km^2$) can become an alternative oxygen source to account for the formation of BIFs and to compensate for the organic carbon burial on the current Earth. The required titania-active area for planetary oxidation and oxygen maintenance in the atmosphere is thus two orders of magnitude larger than the value presented in Table 1, but $f_{titania}$ is still well below than 1. The same holds true for Earth-twin habitable planets around other spectral type stars.

As a result, abiotic oxygen can potentially accumulate in atmospheres of Earth-twin habitable exoplanets around various types of host stars as long as sufficient titania-active areas exist and titania photocatalytic reactions are maintained on their surfaces over a long duration. We therefore conclude that titania can potentially produce abiotic oxygen atmospheres on habitable exoplanets.

**Discussions**

We have presented a hypothesis (the titania hypothesis) that titania may act as an abiotic oxygen source on habitable exoplanets. Several recent works have argued that abiotic oxygen can evolve on habitable exoplanets, especially around M dwarfs due to strong XUV irradiations in upper atmospheres[7-9,21]. The titania photocatalytic mechanism presented in this paper is independent from such photodissociation mechanisms, but share the same view that oxygen may not be a reliable sign of life on habitable exoplanets.

Finally, we note that although the titania hypothesis would not account for the planetary oxidation and the subsequent oxygen atmosphere of the Earth, the titania photocatalytic reactions might have occurred locally on the Earth. One possibility is that this mechanism might have played a role in the formation of BIFs during the early-middle Archean, although such a discussion is beyond the scope of this paper.

## Method

・ Converting NUV flux density (W $m^{-2}$) into oxygen production rate (g $m^{-2}$ $yr^{-1}$)

We derive the order-of-magnitude estimation of oxygen production rate in the Result section as follows:

First, using relational expressions 1 W = 1 J $s^{-1}$ and 1 J = $6.24 * 10^{18}$ eV, NUV

flux density of 1 W $m^{-2}$ corresponds 6.24 * $10^{18}$ eV $m^{-2}$ $s^{-1}$. To convert the energy (eV) into the number of NUV photons shorter than 400 nm, we adopt 3.45 eV per photon (corresponding to 360 nm) as a representative value. This is reasonable since the median wavelength for numbers of NUV (280-400 nm) photon is located around there. Although this treatment is only correct with about 10 percent accuracy, a difference of 10 percent in the number of NUV photons is negligible for the current order-of-magnitude estimation. Thus NUV flux density of 1 W $m^{-2}$ corresponds to 1.81 * $10^{18}$ NUV photons $m^{-2}$ $s^{-1}$.

Next, we adopt the quantum efficiency of 10 percent for the photocatalytic mechanism of titania based on the experimental result[17]. It means that 1 NUV photon produces 0.025 oxygen molecule based on Eqs. (1-2). We use Avogadro's number (6.02 * $10^{23}$) to convert the number of produced oxygen into moles, resulting in 7.5 * $10^{-8}$ mol $m^{-2}$ $s^{-1}$. As 1 mol of oxygen is equivalent to 32 g, the oxygen production rate corresponds 2.4 * $10^{-6}$ g $m^{-2}$ $s^{-1}$. Converting $s^{-1}$ into $yr^{-1}$ using 1 yr = 3.16 * $10^{7}$ s, we obtain the relation 1 W $m^{-2}$ = 76 g $m^{-2}$ $yr^{-1}$ (or, 2.4 mol $m^{-2}$ $yr^{-1}$).

・ Calculating $NUV_{ratio}$ for different types of host stars

First, we integrate 280-400 nm of synthetic stellar spectra using the Phoenix BT-NextGen model[22]. We adopt the effective temperature $T_{eff}$, the surface gravity log g, the stellar radius $R_s$ as summarized in Table 1 to make the synthetic stellar spectra. We assume the bolometric luminosity of host stars $L_s$ using typical values from previous literature (M0 and M6 from Kaltenegger & Traub 2009[23], K2 from Domagal-Goldman et al. 2014[21], F6 from Bruntt et al. 2010[24]). We consider the cases that planets are orbiting at the distance where the effective stellar flux is equal to that for the Sun-Earth system ($S_{eff}$ = 1). The orbital distance $a_p$ can be estimated by the root of $L_s$. We calculate $NUV_{top}$ for various types of host stars by dividing the integrated stellar spectra by $a_p^2$, and compute $NUV_{top,Earth}$ with the same procedure for G2 type star. Finally, $NUV_{ratio}$ is calculated by $NUV_{top}$ / $NUV_{top,Earth}$.

We note that the above estimations for $NUV_{ratio}$ are very rough and may be significantly underestimated for M0 and M6 type cases, since the Phoenix model does not include non-thermal emissions due to stellar activities such as flares. A number of M dwarfs are known to show stronger NUV emissions than the synthetic models[25]. Nevertheless, our main conclusion does not change as stronger NUV flux would encourage the photocatalytic mechanism of titania.

Acknowledgement

This research was supported by NINS Program for Cross-Disciplinary Study. We are grateful for the Monitoring Network for Ultraviolet Radiation, which is operated by the Center for Global Environmental Research (CGER), Natural Institute for Environmental Studies (NIES). N. N. acknowledges supports from NAOJ Fellowship, Inoue Science Research Award, JSPS KAKENHI of Grant-in-Aid for Scientific Research (A) (No. 25247026). S. M. acknowledges supports from JSPS KAKENHI of Grants-in-Aid for Young Scientists (A) (No. 25708011), Challenging Exploratory Research (No.26620160), and Scientific Research on Innovative Areas "An Apple" (No. 25107526).

Author contributions

N.N., T.E., and S.M. conceived this study. N.N. gathered the data from the Monitoring Network for Ultraviolet Radiation. N.N. and N.K. made calculations. N.N., T.E., S.M., and N.K wrote and reviewed the manuscript.

Competing financial interests:

The authors declare no competing financial interests.

Table Caption

Table 1 | Summary of adopted stellar and planetary parameters as well as corresponding $NUV_{ratio}$, $f_{titania,1Gyr}$, and $A_{titania,1Gyr}$. Type means the stellar spectral type. $T_{eff}$ is the effective temperature, log g is the surface gravity, $R_s$ is the stellar radius, and $L_s$ is the bolometric luminosity of host stars. $a_p$ is the semi-major axis of a habitable planet receiving equal effective stellar flux to the case for Sun-Earth system ($S_{eff}$ = 1). $NUV_{ratio}$, $f_{titania,1Gyr}$, and $A_{titania,1Gyr}$ are explained in the body text.

Table 1

| Type | $T_{eff}$ (K) | log g | $R_s$ ($R_{Sun}$) | $L_s$ ($L_{Sun}$) | $a_p$ (AU) | $NUV_{ratio}$ | $f_{titania,1Gyr}$ | $A_{titania,1Gyr}$ ($km^2$) |
|---|---|---|---|---|---|---|---|---|
| M6 | 3000 | 4.5 | 0.15 | 0.0009 | 0.03 | 0.017 | $1*10^{-4}$ | $6*10^{4}$ |
| M0 | 3800 | 4.5 | 0.50 | 0.072 | 0.27 | 0.030 | $7*10^{-5}$ | $3*10^{4}$ |
| K2 | 5000 | 4.5 | 0.73 | 0.33 | 0.58 | 0.31 | $7*10^{-6}$ | $3*10^{3}$ |
| F6 | 6300 | 4.5 | 1.5 | 3.0 | 1.73 | 1.6 | $1*10^{-6}$ | $6*10^{2}$ |
| G2 | 5800 | 4.5 | 1.0 | 1.0 | 1.0 | 1.0 | $2*10^{-6}$ | $1*10^{3}$ |

Figure Captions

Figure 1 | Schematic illustration of the photocatalytic oxygen formation on titania photocatalyst in the presence of water and an electron acceptor. Photon absorption of titania affords an electron ($e^-$) – hole ($h^+$) pair on the photocatalyst. The electron reduces the acceptor and the hole oxidizes water to oxygen on the surface of the titania photocatalyst.

Figure 2 | NUV and total solar flux density data taken at the Hateruma Observatory. The upper panel (a) shows data taken on a clear day (23 June 2013) and the lower panel (b) does the same for a cloudy day (20 June 2013). The horizontal axis indicates time in Japanese Standard Time (Universal Time + 9 hr). The violet line (left axis) plots NUV flux density and the yellow line (right axis) does total solar flux density. The data were obtained in one-minute intervals.

Figure 1

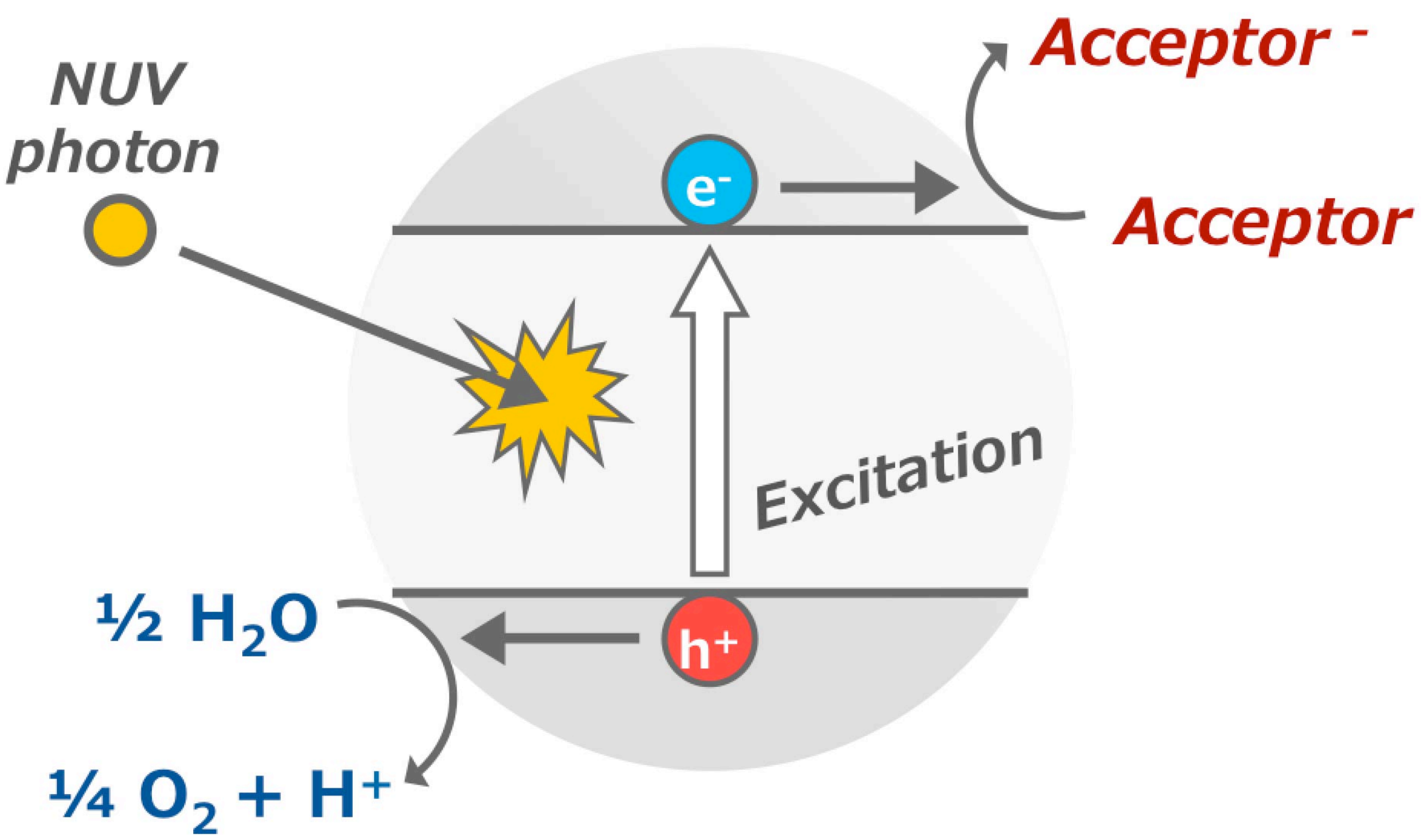

Figure 2

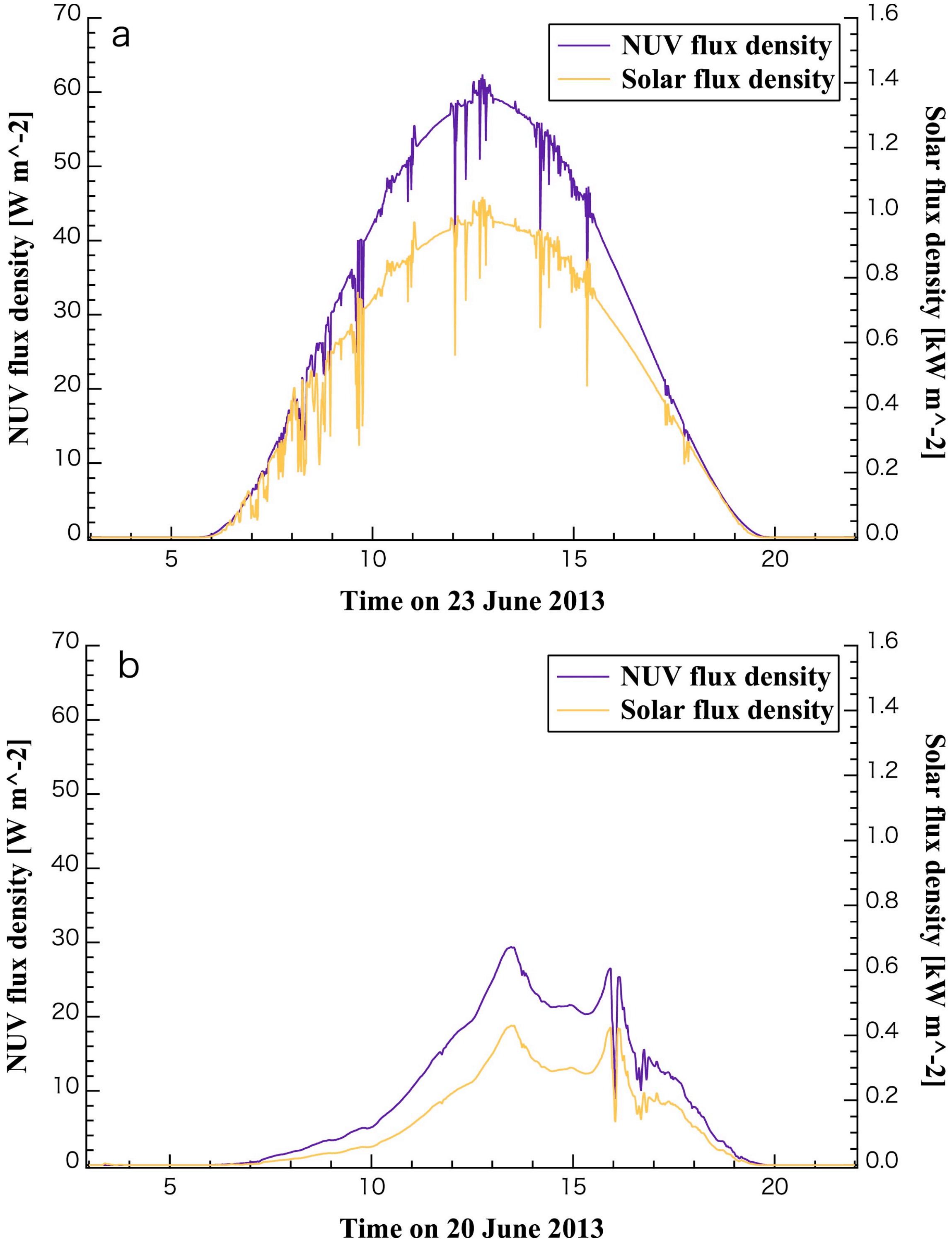